**The Sensitivity of the Magnetic Properties of the $ZnCr_2O_4$ and $MgCr_2O_4$ Spinels to Non-stoichiometry.**

S. E. Dutton[*§], Q. Huang[†], O. Tchernyshyov[‡], C. L. Broholm[‡] and R. J. Cava[*]

[*]Department of Chemistry, Princeton University, Princeton, NJ 08544, USA

[†]Center for Neutron Research, NIST, Gaithersburg, Maryland 20899, USA

[‡]Department of Physics and Astronomy, The Johns Hopkins University, Baltimore, MD 21218, USA

[§] Corresponding author.

ABSTRACT

We report that small amounts of metal atom non-stoichiometry are possible in the $ZnCr_2O_4$ and $MgCr_2O_4$ spinels. The non-stoichiometry, though less than 2%, significantly impacts $T_N$ and the nature of the magnetic correlations above $T_N$. The $Zn_{1+x}Cr_{2-x}O_4$ spinel is particularly sensitive. While stoichiometric $ZnCr_2O_4$ displays antiferromagnetic short range correlations in the susceptibility above $T_N$, ferromagnetic correlations are observed in non-stoichiometric, hole doped $Zn_{1+x}Cr_{2-x}O_4$. The $Mg_{1+x}Cr_{2-x}O_4$ spinels are less profoundly affected by non-stoichiometry, though significant changes are also observed. We contrast the magnetic properties of $Zn_{1+x}Cr_{2-x}O_4$ and $Mg_{1+x}Cr_{2-x}O_4$ (x=0, 0.02, 0.04) with those of materials with the equivalent amounts of isovalent non-magnetic $Ga^{3+}$ substituted on the $Cr^{3+}$ site to separate the effects of static site disorder and hole doping.

## 1. INTRODUCTION

The $ZnCr_2O_4$, $MgCr_2O_4$ and $CdCr_2O_4$ spinels are among the most extensively studied geometrically frustrated magnetic systems.[1-12] Despite having a relatively simple crystal structure (FIG. 1 inset), with the $Cr^{3+}$ ions forming a 3D array of corner sharing tetrahedra, the low temperature magnetostructural behavior is complex. All of the chromium(III) spinels undergo an antiferromagnetic ordering transition with $T_N < 13$ K, accompanied by a structural transition to lower crystallographic symmetry. Given the presence of isotropic spin $Cr^{3+}$ ($d^3$), the lowering of symmetry is thought to be driven by spin-lattice coupling in a spin-Peierls-like transition.[8] Both the structural and magnetic behavior at low temperature are highly dependent on the A ion, as are the short range magnetic correlations observed above $T_N$. In $ZnCr_2O_4$, the most extensively studied of the family, significant differences in properties have been observed[7-9], but no connection has been made to the chemistry of the compound. In addition, the exact nature of the structural and magnetic ordering in $ZnCr_2O_4$ has yet to be resolved; the magnetic and structural transitions being much more complex than for $MgCr_2O_4$ and $CdCr_2O_4$. Here we report that percent level metal atom non-stoichiometry, strongly related to the synthetic conditions, can be present in both the $ZnCr_2O_4$ and $MgCr_2O_4$ spinels. In these non-stoichiometric $A_{1+x}Cr_{2-x}O_4$ compositions, excess non-magnetic Zn or Mg is found in the Cr sites in addition to the usual A sites. To maintain charge neutrality, a corresponding fraction of $Cr^{3+}$ is oxidized to $Cr^{4+}$. Thus non-stoichiometry induces both static disorder, from the presence of Zn or Mg on the Cr lattice, and dynamic disorder, through hole doping. In the case of $ZnCr_2O_4$, the magnetic properties are profoundly affected, and comparison with Ga substituted analogues, which introduce static disorder, shows that it is the hole doping, not the disorder, that has the largest impact on the magnetism.

## 2. EXPERIMENTAL

Polycrystalline samples of $A_{1+x}Cr_{2-x}O_4$ and $ACr_{2-x}Ga_xO_4$ (A=Zn, Mg x=0,0.02,0.04) were prepared using solid-state synthesis from high purity binary oxide starting reagents: zinc (II) oxide (Alfa Aeaser 99.99 %), magnesium (II) oxide (Alfa Aeaser 99.95 %), chromium (III) oxide (Fisher Scientific) and pre-dried gallium (III) oxide (Johnson Matthey 99.999%). The appropriate quantities of reagents were intimately mixed and pressed into pellets prior to heating at 800 $^{o}$C for 12 hours, followed by an additional firing at 1200 $^{o}$C for 24 hours. This high temperature reaction step was repeated until a phase pure product, as determined by powder X-ray diffraction, was obtained. For samples of $ZnCr_2O_4$ and $ACr_{2-x}Ga_xO_4$ the synthesis was carried out under flowing Ar gas. Synthesis of all other compositions (including $MgCr_2O_4$) was carried out in air. Samples of $ZnGa_2O_4$ and $MgGa_2O_4$ were synthesized from stoichiometric mixtures of the binary oxides heated to 1100 $^{o}$C for 24 hours in air.

X-ray powder diffraction patterns were collected on a Bruker D8 Focus operating with Cu K$\alpha$ radiation and a graphite diffracted beam monochromator. High resolution data for structural refinement were collected over the angular range $5 \leq 2\theta \leq 90^{o}$ with a step size of $\Delta 2\theta = 0.04^{o}$. Rietveld refinement[13] of the structures was carried out using the GSAS suite of programs[14] with the EXPGUI interface.[15] Backgrounds were fitted using a Chebyshev polynomial of the first kind and the peak shape was modeled using a pseudo-Voigt function.

Magnetic susceptibility and specific heat measurements were made using a Quantum Design Physical Properties Measurement System (PPMS) and a Quantum Design MPMS-5 Superconducting Quantum Interference Device (SQUID) magnetometer. Temperature dependent magnetization measurements of finely ground samples were collected under 0.1 and 1 T fields between 2 and 300 K after cooling in zero field. In selected samples, temperature dependent magnetization measurements were also collected between 5 and 30 K after cooling in the measuring field (1 T). The specific heat of all samples was measured in zero field over a narrow temperature range spanning $T_N$, $5 \leq T \leq 15$ K. The lattice contribution to the specific heat was deducted by subtracting the specific heat of the isostructural non-magnetic analogues, $ZnGa_2O_4$ and $MgGa_2O_4$; since no change in the symmetry of either of these phases is observed on cooling, the resultant specific heat, $C_{magst}$, includes contributions from both the magnetic and structural transitions near 12 K. Pellets for heat capacity for the Mg spinels and $ZnCr_2O_4$ were prepared by heating for 12 hours at 1250 $^{o}$C in the same gas as for the initial synthesis. Specific heat measurements of non-stoichiometric and gallium doped $ZnCr_2O_4$ were made using pressed pellets of the sample plus silver powder. The specific heat of silver was subsequently deducted.[16] Sintered pellets of the non-magnetic analogues, $ZnGa_2O_4$ and $MgGa_2O_4$, were prepared by heating to 1150 $^{o}$C for 12 hours in air.

Neutron powder diffraction patterns were collected on $ZnCr_2O_4$ and $Zn_{1.04}Cr_{1.96}O_4$ at the NIST Center for Neutron Research on the high-resolution powder neutron diffractometer (BT-1) using neutrons with wavelength $\lambda$=1.5403 Å from a Cu(311) monochromator. Collimators with horizontal divergences of 15°, 20° and 7° of arc were used before and after the monochromator and after the sample, respectively. Data were collected over the angular range $3 \leq 2\theta \leq 168°$ with a step size of $\Delta 2\theta$=0.05°. Samples were placed in vanadium cans ($\phi$=10 mm) and patterns were collected above $T_N$ at 15 K. The intensity of selected magnetic and nuclear peaks as a function of temperature through the magnetic and structural transition was also measured. Structural refinement using the Rietveld method[13] was carried out using the Fullprof suite of programs.[17] Backgrounds were fitted using a refined interpolation of data points and the peak shape modeled using a pseudo-Voigt function.

## 3. RESULTS

For both the $MgCr_2O_4$ and $ZnCr_2O_4$ systems, small amounts of metal atom non-stoichiometry are possible. At the limiting composition, $A_{1.04}Cr_{1.96}O_4$, in addition to the structural disorder introduced by $A^{2+}$ cations on the Cr network, 2% of the $Cr^{3+}$ is oxidized to $Cr^{4+}$. For both A cations, the hole doping due to the non-stoichiometry resulted in a change in color from green in the stoichiometric and Ga-doped samples to brown/black in the samples containing $Cr^{4+}$. What appeared to be "stoichiometric" $ZnCr_2O_4$ could be prepared in air, but, despite the absence of detectable impurity peaks in the X-ray diffraction pattern, the brownish color and magnetic properties of the air-synthesized samples were characteristic of hole doped non-stoichiometric $Zn_{1+x}Cr_{2-x}O_4$; only in samples prepared under an Ar atmosphere could green, stoichiometric $ZnCr_2O_4$ be prepared. In neither the non-stoichiometric nor the Ga-doped samples are distortions from the cubic spinel structure observed at room temperature; details from the X-ray refinements are given in TABLES I and II. The changes to the structure on doping are small, as would be expected for such small deviations from the parent compound, and, relative to the range of lattice parameters previously reported for $ZnCr_2O_4$ (a=8.308-8.330 Å[4, 7, 12, 18]) are insignificant. Comparison of the structural parameters obtained by refinement of the neutron diffraction data for $Zn_{1.04}Cr_{1.96}O_4$ and $ZnCr_2O_4$ collected at 15 K (TABLE III and FIG. 1) shows only a small difference of the oxygen position in the two phases. This reflects a small contraction of the average $(Cr_{0.98}Zn_{0.02})O_6$ octahedral site in $Zn_{1.04}Cr_{1.96}O_4$. No evidence for Ga/Zn or Ga/Mg mixing on the A site is observable in diffraction data from the Ga doped samples. This type of disorder is not expected in the $ZnCr_2O_4$ case because both $ZnCr_2O_4$ and $ZnGa_2O_4$[19] are normal spinels, but may be possible in the $MgCr_2O_4$ case because $MgGa_2O_4$[20] is an inverse spinel. However, in the current study such disorder, if present, is not relevant, because neither the Cr valence nor the degree of disorder on the $Cr^{3+}$ magnetic lattice would be impacted.

The temperature dependent magnetic susceptibility, $\chi$=dM/dT, in the $Zn_{1+x}Cr_{2-x}O_4$ spinel changes significantly (FIG. 2) with non-stoichiometry. All samples show an antiferromagnetic transition, $T_N$, defined by a maximum in $d(\chi T)/dT$, near 12 K (TABLE I). No irreversibility between the ZFC and FC magnetic susceptibility in either of the limiting compositions is observed, and thus this small level of non-stoichiometry and/or disorder does not inhibit long range magnetic ordering. A decrease in $T_N$ is observed in both the hole and Ga-doped samples. This effect is most dramatic in the non-stoichiometric hole doped samples, where a shift of 1.8 K relative to stoichiometric $ZnCr_2O_4$ ($T_N$=12.6 K) is observed in $Zn_{1.04}Cr_{1.96}O_4$ ($T_N$=10.8 K). The equivalent Ga substituted compound, $ZnCr_{1.96}Ga_{0.04}O_4$, in which only structural disorder is introduced, shows a much smaller and barely significant decrease in $T_N$, to 12.4 K. At temperatures just above $T_N$ the magnetization when compared to $ZnCr_2O_4$ is larger in all of the doped samples. This effect is more pronounced for the non-stoichiometric materials. In non-stoichiometric $Zn_{1+x}Cr_{2-x}O_4$ the shape of the feature at $T_N$ changes, showing a large increase in magnetization just above the transition and a sharper more pronounced decrease in $\chi$ at $T_N$. At high temperature, $150\leq T\leq 300$ K, fits to the Curie-Weiss law, $\chi-\chi_o=C/T-\theta$, were carried out and are detailed in TABLE I. The changes on doping are relatively small, though in all the doped samples a small decrease in the Weiss temperature $\theta$ is observed. No significant changes to the effective moment per Cr, $\mu_{eff}$, are observed on doping, as is expected given the small percentage of $Cr^{4+}$ introduced by non-stoichiometry.

The changes in the magnetic susceptibility of non-stoichiometric and doped $MgCr_2O_4$ are less dramatic (FIG. 3) than for $ZnCr_2O_4$. The transition temperature, $T_N$, is observed to be highly dependent on the composition (TABLE II). Opposite to the shift observed in $T_N$ for the Zn spinels, non-stoichiometry results in a smaller decrease in $T_N$ when compared to the Ga-substitution-induced structural disorder. In the case of $Mg_{1.02}Cr_{1.98}O_4$ no significant deviation from $T_N$ in stoichiometric $MgCr_2O_4$ is observed. As for the Zn analogues, the absence of hysteresis between the ZFC and FC susceptibility indicates the presence of long range ordering. Around $T_N$ the susceptibilities for the non-stoichiometric, hole doped, and disordered Ga-doped samples are very similar, the exception being the most non-stoichiometric material $Mg_{1.04}Cr_{1.96}O_4$, where the deviations from Curie-Weiss behavior above $T_N$ are reminiscent of those observed in $Zn_{1+x}Cr_{2-x}O_4$. The similarities between non-stoichiometric and Ga-doped samples are also seen in the magnetic parameters obtained from fitting to the Curie-Weiss law for $150 \leq T \leq 300$ K.

By rearranging the Curie-Weiss law to give, for an antiferromagnet, $C/\chi|\theta| = T/|\theta|+1$, it can be seen that in a plot of the normalized inverse susceptibility, $C/\chi|\theta|$, as a function of normalized temperature, $T/|\theta|$, perfect Curie-Weiss behavior would result in a slope of gradient 1 intersecting the y axis at 1. Such plots have been used in comparing the behavior of geometrically frustrated magnets[21-23], allowing differences in the nature of correlations above $T_N$ to be identified and as a meaningful way to compare magnetic susceptibilities with significant differences in magnitude. Normalized inverse susceptibility plots are shown for the doped and non-stoichiometric $ZnCr_2O_4$ and $MgCr_2O_4$ systems in FIG. 4. These plots show that the magnetic susceptibility is dependent on both the type of dopant and the A ion. The large degree of frustration in all of the spinels can be seen in the suppression of the ordering temperature, $T_N/|\theta| \sim 0.03$ (typically for a frustrated system $T_N/|\theta| < 0.1$). In stoichiometric $ZnCr_2O_4$, antiferromagnetic short range correlations, i.e. positive deviations from Curie-Weiss behavior, are observed above $T_N$. However, ferromagnetic short range correlations, i.e. negative deviations from Curie-Weiss behavior, are observed in the non-stoichiometric hole doped compositions. The Ga-doped samples of $ZnCr_2O_4$ exhibit stronger antiferromagnetic short range correlations than the stoichiometric phase. Comparison of the normalized susceptibility for $Zn_{1.04}Cr_{1.96}O_4$ and $ZnCr_{1.96}Ga_{0.04}O_4$ shows a dramatic difference: for static disorder on the Cr lattice, enhanced antiferromagnetic correlations are induced above $T_N$, whilst disorder and hole doping enhances the ferromagnetic correlations. Curiously, one can infer from FIG. 4 that for non-stoichiometry near $Zn_{1.04}Cr_{1.96}O_4$, the induced FM and intrinsic AFM short range correlations compensate for one another, giving apparently ideal Curie-Weiss-like behavior to lower $T/|\theta|$ than for the stoichiometric material. In the $MgCr_2O_4$ system both non-stoichiometric and doped compositions display positive deviations from the Curie-Weiss law above $T_N$ indicative of antiferromagnetic short range correlations. The onset temperature of the correlations appears to be independent of whether the dopant induces disorder or both disorder and hole doping.

Specific heat measurements are shown in FIGS. 5 and 6. The specific heat allows the changes in entropy resulting from the structural and magnetic transitions to be probed. For an antiferromagnet, changes to the magnetic entropy, as reflected in the specific heat, have similar behavior to $d(\chi T)/dT$.[24] On consideration of the specific heat it can be seen that in both the doped $ZnCr_2O_4$ and $MgCr_2O_4$ systems the peaks in the specific heat and the magnetic entropy, $d(\chi T)/dT$¸ are coincident. This indicates that the structural and magnetic transitions occur simultaneously, as some part of the specific heat anomaly must be due to the structural phase transition.

In non-stoichiometric $MgCr_2O_4$ the decrease in the temperature of the transitions is small, and as the non-stoichiometry increases a broadening in the peak in the specific heat is observed. A larger shift in the transition temperature is observed in the Ga-doped samples; from $T_N$=12.4 K in stoichiometric $MgCr_2O_4$ to $T_N$=10.5 K in $MgCr_{1.96}Ga_{0.04}O_4$. In contrast, in non-stoichiometric $Zn_{1+x}Cr_{2-x}O_4$ (FIG. 5) the maxima in the specific heat shift to dramatically lower temperatures as the non-stoichiometry increases. A smaller decrease in temperature is observed in the Ga-doped samples. For both non-stoichiometric and Ga-doped $ZnCr_2O_4$ the broadening of the transitions observed in specific heat and the magnetic entropy indicate a more diffuse magnetic transition relative to stoichiometric $ZnCr_2O_4$ and the non-stoichiometric or doped $MgCr_2O_4$ systems. The origin of the more diffuse transition is not currently known, however if the phase transition indeed results from small structural changes that break the symmetry of otherwise frustrating interactions,[6] it is plausible that small changes in stoichiometry or disorder impact the detailed character of the transitions significantly.

To investigate the changes in the structure and magnetic ordering in $Zn_{1-x}Cr_{2-x}O_4$ at $\sim T_N$ in more detail, low temperature neutron diffraction data were collected on $Zn_{1.04}Cr_{1.96}O_4$ and $ZnCr_2O_4$. On cooling below $T_N$, magnetic peaks appear in the diffraction patterns, and a broadening of the $(800)_N$ nuclear reflection, indicative of a reduction in structural symmetry, is observed. The structural and magnetic transitions were investigated by following the evolution of peaks that show a change in intensity associated only with the individual transitions (FIG. 7); for the nuclear transition this is the $(800)_N$ reflection at $2\theta \sim 96^o$ and for the magnetic transition a peak indexed as $(\frac{1}{2}\frac{1}{2}1)_M$ at $2\theta \sim 22.6^o$ using the supercell proposed by Ji *et al.*[6] As in the magnetic susceptibility and specific heat a clear decrease in the temperature of the structural and magnetic transitions is observed in $Zn_{1.04}Cr_{1.96}O_4$ compared to $ZnCr_2O_4$. For both compositions, the onset of the structural and magnetic transitions is the same within the precision of the measurement. Thus the structural and magnetic transitions are coupled in both the stoichiometric and the non-stoichiometric compounds. Since no irreversibility between the ZFC and FC susceptibility is observed in either $ZnCr_2O_4$ or $Zn_{1.04}Cr_{1.96}O_4$ the hysteresis observed in the evolution of the structural and magnetic transitions at $T_N$ from neutron diffraction may be a result of the finite thermal sweep rate during the diffraction measurements.

## 4. DISCUSSION AND CONCLUSION

Non-stoichiometry and the resultant hole doping in $ZnCr_2O_4$ and $MgCr_2O_4$ spinels has a profound effect on the magnetic properties even at a doping level of 2% or lower. The low temperature behavior is highly dependent on both the A ions and the type of dopant. In non-stoichiometric $Zn_{1+x}Cr_{2-x}O_4$ the changes observed are complex. The change in the nature of the correlations above $T_N$ in the non-stoichiometric hole doped samples is the most significant indication of the effect of hole doping on the magnetic properties. These correlations above $T_N$ are indicative of short range ordering and, due to the mangetostructural coupling, are expected to play a role in the formation of both the crystallographic and magnetic long range ordered states below $T_N$. The presence of short range ferromagnetic correlations above $T_N$ in non-stoichiometric $Zn_{1+x}Cr_{2-x}O_4$ may therefore be indicative of a change in the magnetic or/and structural ordering at lower temperatures. $MgCr_2O_4$ appears to be more robust, with smaller though significant changes in behavior induced by non-stoichiometry. This is consistent with the simpler structural and magnetic ordering transition observed in stoichiometric $MgCr_2O_4$. We have not studied the structure of doped or non-stoichiometric $MgCr_2O_4$ below $T_N$, but the relatively small systematic changes to the magnetization and the specific heat suggest that the magnetic and structural characteristics are similar to that of stoichiometric $MgCr_2O_4$. The coincidence of the peaks in the specific heat and

d($\chi$T)/dT indicates that the structural and magnetic transitions occur simultaneously in both $MgCr_2O_4$ and $ZnCr_2O_4$, and thus small levels of doping or non-stoichiometry are not sufficient to disrupt the coupling of the structural and magnetic transitions. Thus, as for the stoichiometric case, it can be inferred that in doped $MgCr_2O_4$ and $ZnCr_2O_4$ spin-lattice coupling drives the concurrent structural and magnetic transitions at low temperature. This single transition occurs at a temperature where the magnetic interactions overcome the stiffness of the lattice; at that temperature a lattice distortion sets in, decreasing the symmetry, relieving the magnetic frustration, and allowing magnetic ordering. In the non-stoichiometric and Ga-doped systems both the structure and the magnetic network are disrupted by the presence of the small number of "impurity atoms" and it may be that it is through impacting the lattice stiffness that hole doping, non-stoichiometry and disorder change $T_N$. Rationalizing the changes in $T_N$ thus requires comprehending in more detail the changes in both the structural and magnetic regimes introduced by the introduction of Ga or non-stoichiometry.

Because the sizes of the magnetic lattices in $MgCr_2O_4$ and $ZnCr_2O_4$ are very similar, the difference in magnetic properties of $MgCr_2O_4$ and $ZnCr_2O_4$ cannot simply be ascribed to differences in crystal structure above $T_N$. Instead we speculate that they may be due to the subtle differences in bonding, and subsequently the magnetic exchange interactions, that are a consequence of the difference in electronegativity between Zn and Mg. Given the difference in the behavior of the parent compounds, it is not surprising that $ZnCr_2O_4$ and $MgCr_2O_4$ exhibit different properties when doped or non-stoichiometric. The divergence in the behavior of the doped $ZnCr_2O_4$ and $MgCr_2O_4$ spinels highlights the difficulties in generalizing the relationships between chemical perturbations to the magnetic lattice and the magnetic properties in geometrically frustrated magnets. In addition to the chromium(III) spinels the only other hole doped geometrically frustrated system containing $Cr^{3+}$ reported is $\beta$-$CaCr_2O_4$.[21] In all three systems, hole doping induces different changes in the magnetism compared to static doping; the behavior of each individual system is also different. Thus charge doping provides an additional means of manipulating the sometimes delicate balance of interactions within geometrically frustrated magnetic systems.

## 4. ACKNOWLEDGEMENTS

The authors wish to thank Shuang Jia and E. Climent-Pascual for their helpful discussions. This research was supported by the US Department of Energy, Division of Basic Energy Sciences, Grant DE-FG02-08ER46544.

TABLE I. Structural and Magnetic parameters for non-stoichiometric and Ga doped $ZnCr_2O_4$

| | $a$ / Å | O position / x | $T_N$ / K | C / emu $Oe^{-1}$ $mol_{Cr}^{-1}$ $K^{-1}$ | $\mu_{eff}$ / $\mu_B$ per Cr | θ / K |
|---|---|---|---|---|---|---|
| $ZnCr_2O_4$ | 8.31948(14) | 0.2622(2) | 12.6(1) | 1.83 | 3.82 | -400 |
| $Zn_{1.02}Cr_{1.98}O_4$ | 8.32166(10) | 0.2623(2) | 11.0(1) | 1.91 | 3.91 | -373 |
| $Zn_{1.04}Cr_{1.96}O_4$ | 8.32055(13) | 0.2635(3) | 10.8(1) | 1.92 | 3.91 | -372 |
| $ZnCr_{1.98}Ga_{0.02}O_4$ | 8.3180(2) | 0.2613(4) | 12.4(1) | 1.88 | 3.88 | -374 |
| $ZnCr_{1.96}Ga_{0.04}O_4$ | 8.3194(4) | 0.2599(4) | 12.4(1) | 1.95 | 3.95 | -386 |

TABLE II. Structural and Magnetic parameters for non-stoichiometric and Ga doped $MgCr_2O_4$

| | $a$ / Å | O position / x | $T_N$ / K | C / emu $Oe^{-1}$ $mol_{Cr}^{-1}$ $K^{-1}$ | $\mu_{eff}$ / $\mu_B$ per Cr | θ / K |
|---|---|---|---|---|---|---|
| $MgCr_2O_4$ | 8.32768(11) | 0.2591(3) | 12.8(1) | 1.97 | 3.97 | -433 |
| $Mg_{1.02}Cr_{1.98}O_4$ | 8.32789(12) | 0.2562(2) | 12.9(1) | 1.94 | 3.94 | -419 |
| $Mg_{1.04}Cr_{1.96}O_4$ | 8.32674(11) | 0.2585(2) | 12.4(1) | 1.92 | 3.93 | -409 |
| $MgCr_{1.98}Ga_{0.02}O_4$ | 8.3280(2) | 0.2553(2) | 11.9(1) | 1.91 | 3.91 | -421 |
| $MgCr_{1.96}Ga_{0.04}O_4$ | 8.3263(2) | 0.2553(3) | 10.5(1) | 1.89 | 3.89 | -412 |

TABLE III. Structural parameters and bond lengths of $ZnCr_2O_4$ and $Zn_{1.04}Cr_{1.96}O_4$ at 15 K.

| 15 K | | $ZnCr_2O_4$ | $Zn_{1.04}Cr_{1.96}O_4$ |
|---|---|---|---|
| Space Group | | $Fd\bar{3}m$ | $Fd\bar{3}m$ |
| *a* / Å | | 8.320721(5) | 8.321816(6) |
| $R_f$ | | 1.12 | 1.49 |
| $\chi^2$ | | 1.24 | 1.24 |
| Zn<br>8*a*<br>(⅛,⅛,⅛) | $B_{iso}$ / Å$^2$ | 0.16(2) | 0.20(3) |
| Cr/Zn<br>16*d*<br>(½,½,½) | $B_{iso}$ / Å$^2$ | 0.13(2) | 0.25(3) |
| O<br>32*e*<br>(x,x,x) | x<br>$B_{iso}$ / Å$^2$ | 0.26157(3)<br>0.21(1) | 0.26169(5)<br>0.28(2) |
| Zn-O / Å | | 1.9717(3) | 1.9737(4) |
| Cr/Zn-O / Å | | 1.9921(3) | 1.9914(4) |
| Cr/Zn-Cr/Zn / Å | | 2.9469786(1) | 2.9473624(1) |

FIG. 1. Observed (o) and calculated (-) neutron diffraction data for $Zn_{1.04}Cr_{1.96}O_4$ collected at 15 K. The difference curve is also shown; reflection positions for the cubic spinel are indicated by the vertical lines. A polyhedral model of the $ZnCr_2O_4$ structure is inset; blue and yellow polyhedra represent Zn and Cr respectively.

FIG. 2. Magnetic susceptibility ($\chi$) as a function of temperature for samples of non-stoichiometric and doped $ZnCr_2O_4$, the inset shows the magnetic susceptibility around $T_N$ in more detail. Closed symbols represent non-stoichiometric $Zn_{1+x}Cr_{2-x}O_4$ and open symbols doped $ZnCr_{2-x}Ga_xO_4$.

FIG. 3. Magnetic susceptibility ($\chi$) as a function of temperature for samples of non-stoichiometric and doped $MgCr_2O_4$, the inset shows the magnetic susceptibility around $T_N$ in more detail. Closed symbols represent non-stoichiometric $Mg_{1+x}Cr_{2-x}O_4$ and open symbols doped $MgCr_{2-x}Ga_xO_4$.

FIG. 4. Normalized inverse susceptibility plots for non-stoichiometric and doped $ZnCr_2O_4$ (upper) and $MgCr_2O_4$ (lower).

FIG. 5. Magnetic contribution to the specific heat in non-stoichiometric and doped $ZnCr_2O_4$. The lower panel shows the differential of the magnetic entropy, $d(\chi T)/dT$.

FIG. 6. Magnetic contribution to the specific heat in non-stoichiometric and doped $MgCr_2O_4$. The lower panel shows the differential of the magnetic entropy, $d(\chi T)/dT$.

FIG. 7. Evolution of the nuclear reflection $(800)_N$ (upper) and the magnetic $(\frac{1}{2}\frac{1}{2}1)_M$ reflection (lower) as a function of temperature in $ZnCr_2O_4$ (left) and $Zn_{1.04}Cr_{1.96}O_4$ (right). Data are shown for both warming (black squares) and cooling (green triangles). The solid lines are a guide to the eye.

FIG. 1.

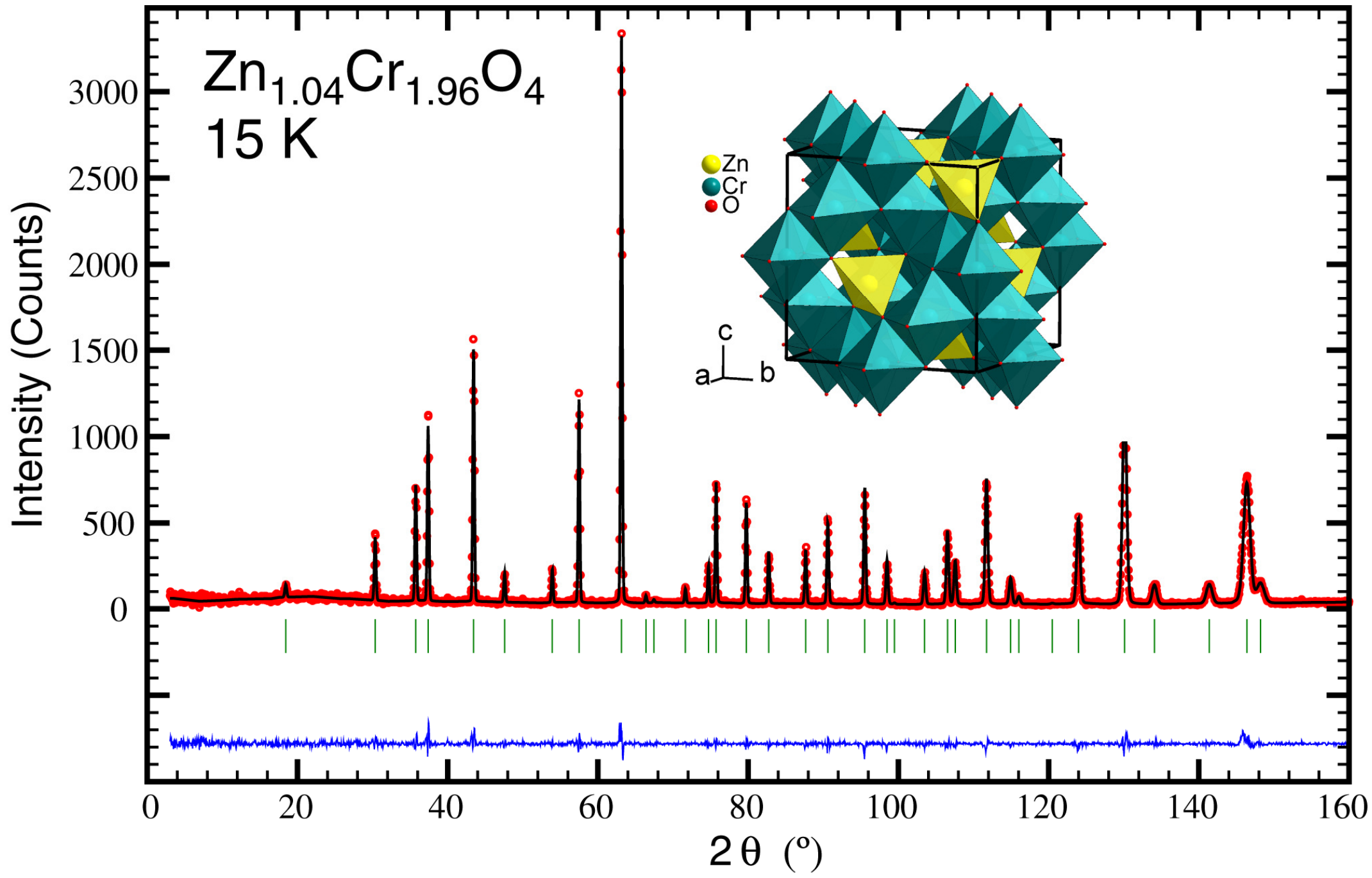

FIG. 2.

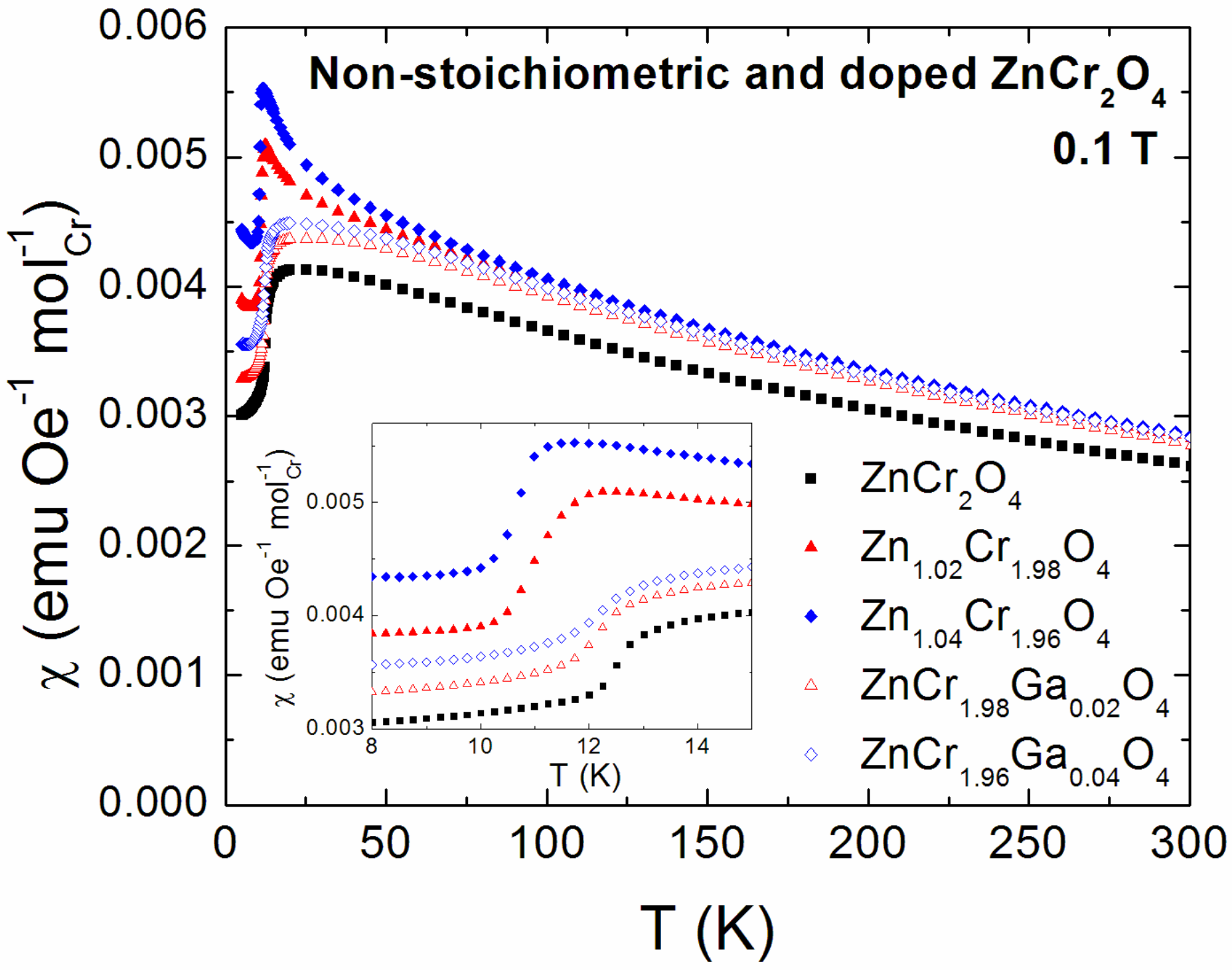

FIG. 3.

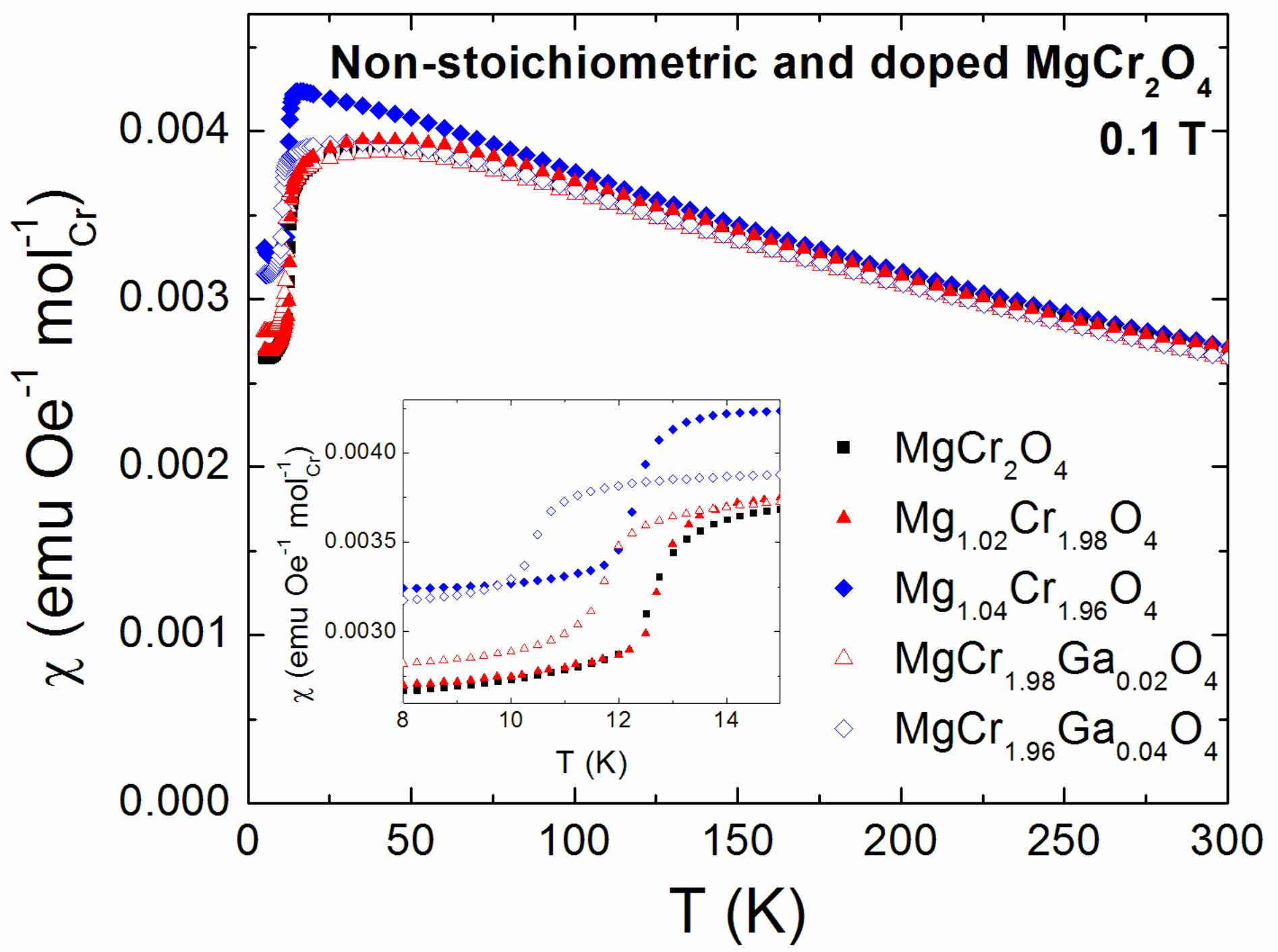

Non-stoichiometric and doped MgCr$_2$O$_4$
0.1 T
χ (emu Oe$^{-1}$ mol$^{-1}_{Cr}$)
T (K)
MgCr$_2$O$_4$
Mg$_{1.02}$Cr$_{1.98}$O$_4$
Mg$_{1.04}$Cr$_{1.96}$O$_4$
MgCr$_{1.98}$Ga$_{0.02}$O$_4$
MgCr$_{1.96}$Ga$_{0.04}$O$_4$

FIG. 4.

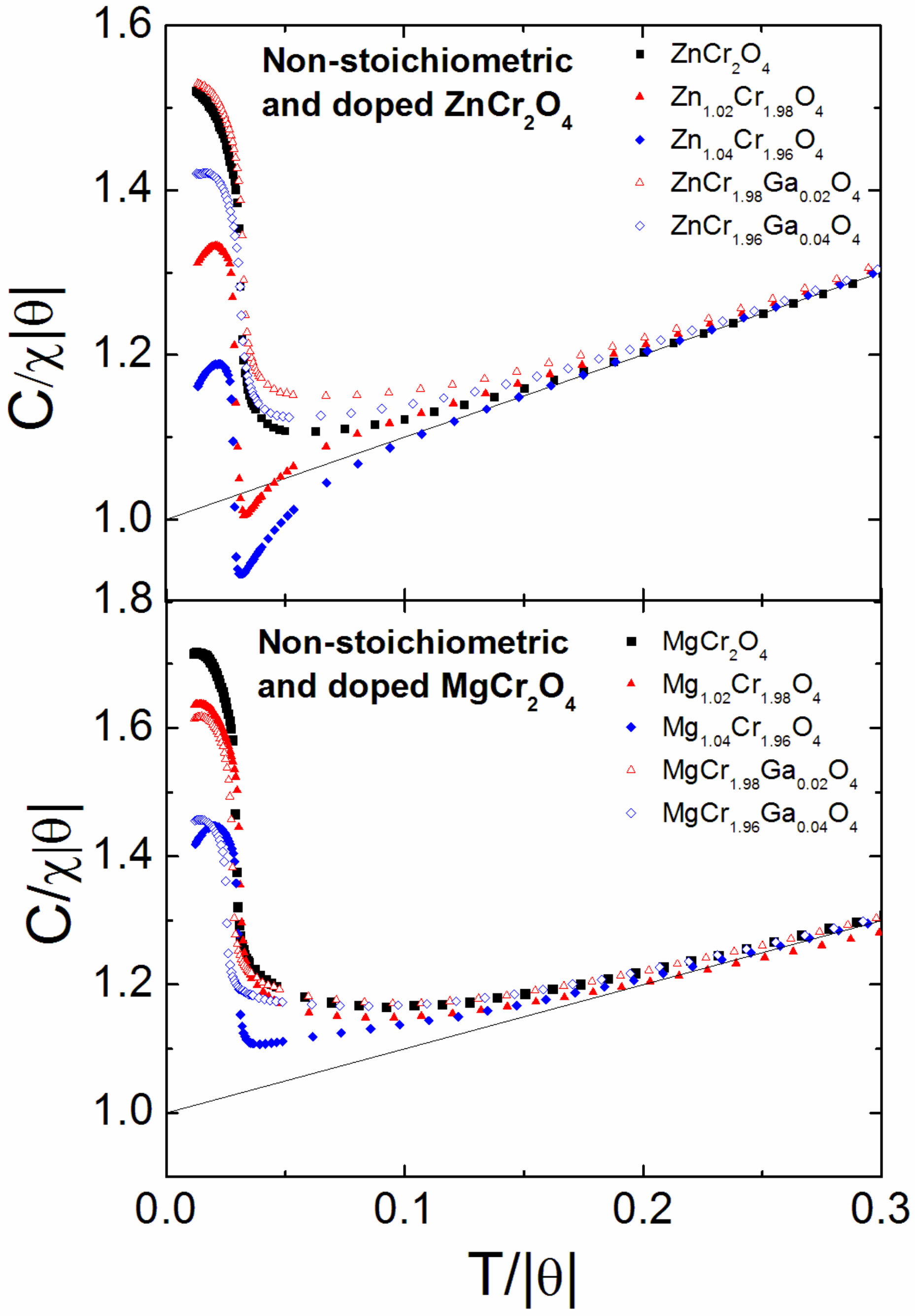

FIG. 5.

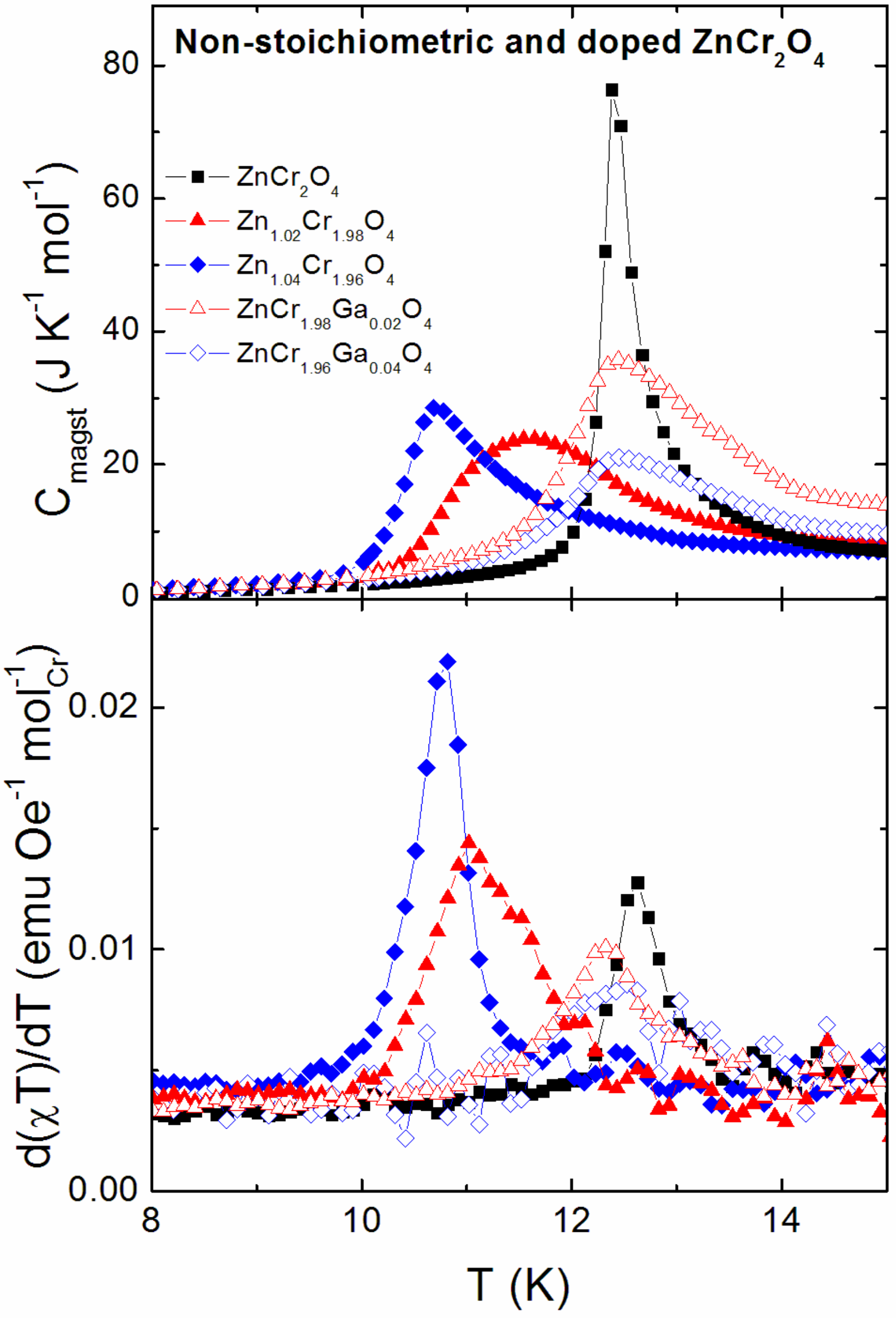

FIG. 6.

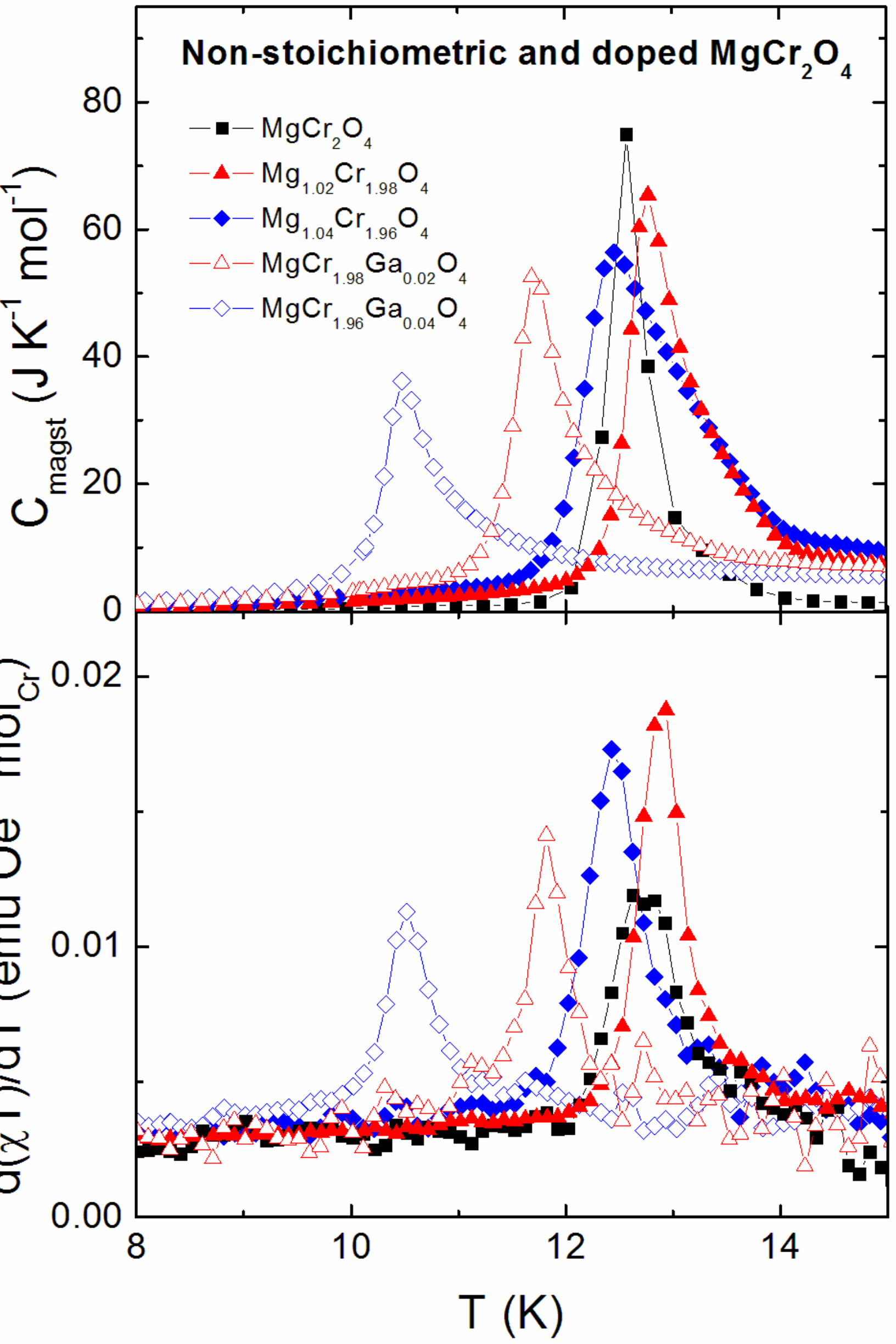

FIG. 7.

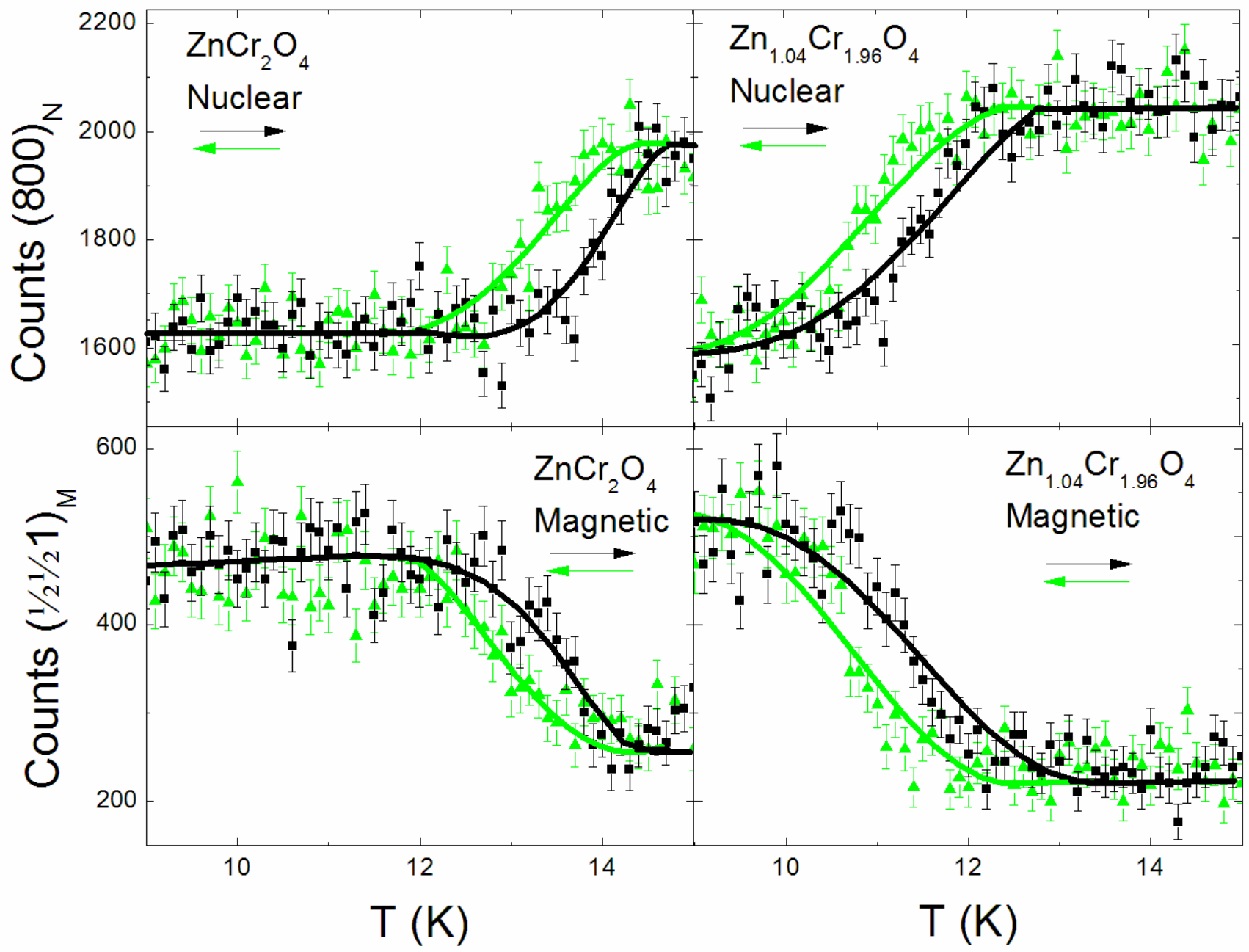